\documentclass[12pt]{article}
\usepackage{a4,epsfig,rotating}

\newcommand{\sqs}{\ensuremath{\sqrt{s}}}
\newcommand{\mh}{\ensuremath{m_{\mathrm H^\pm}}}
\newcommand{\hpm}{\ensuremath{\mathrm H^{\pm}}}

\newcommand{\hh}{\ensuremath{\mathrm{H^+H^-}}}
\newcommand{\hp}{\ensuremath{\mathrm{H^+}}}

\newcommand{\btn}{\mbox{{\cal B}($\mathrm{H^+\!\to\!\tau^+\nu_\tau}$)}}
\newcommand{\bcs}{\mbox{{\cal B}($\mathrm{H^+\!\to\!c\bar{s}}$)}}
\newcommand{\tntn}{\ensuremath{\tau^+\nu_{\tau}\tau^-\bar{\nu}_{\tau}}}
\newcommand{\tncs}{\ensuremath{\mathrm{c\bar{s}}\tau^-\bar{\nu}_{\tau}}}
\newcommand{\cs}{\ensuremath{\mathrm{c\bar{s}}}}
\newcommand{\tn}{\ensuremath{\tau^+\nu_{\tau}}}
\newcommand{\cstn}{\ensuremath{\mathrm{\bar{c}s}\tau^+\nu_{\tau}}}
\newcommand{\cscs}{\ensuremath{\mathrm{c\bar{s}s\bar{c}}}}

\newcommand{\itee}{\ensuremath{e^+e^-}}
\newcommand{\ww}{\ensuremath{\mathrm{W^+W^-}}}
\newcommand{\qq}{\ensuremath{\mathrm{q\bar{q}}}}
\newcommand{\ff}{\ensuremath{\mathrm{f\bar{f}}}}

\newcommand{\gev}{\ensuremath{\mathrm{GeV}}}
\newcommand{\gevc}{\ensuremath{\mathrm{GeV}/c}}
\newcommand{\gevcc}{\ensuremath{\mathrm{GeV}/c^2}}
\newcommand{\invpb}{\ensuremath{\mathrm{pb^{-1}}}}
\newcommand{\sqrs}{\ensuremath{\sqrt{s}}}

\newcommand{\wwemnqq}{\ensuremath{\mathrm{W^+W^- \to (e/\mu) \nu q \bar{q}^\prime}}}
\newcommand{\wwtnqq}{\ensuremath{\mathrm{W^+W^- 
                                 \to \tau \nu q \bar{q}^\prime}}}
\newcommand{\wwlnqq}{\ensuremath{\mathrm{W^+W^- 
                                 \to \ell \nu q \bar{q}^\prime}}}
\newcommand{\eett}{\ensuremath{\mathrm{e^+ e^- \to \tau^+ \tau^-}}}
\def\PLB#1#2#3{{ Phys. Lett. }{\bf B#1 }(#2) #3}

\def\PRD#1#2#3{{ Phys. Rev. }{\bf D#1 }(#2) #3}

\def\EPJ#1#2#3{{ Eur. Phys. J. }{\bf C#1 }(#2) #3}
\def\NIM#1#2#3{{ Nucl. Instrum. and Methods}
{\bf A#1 }(#2) #3}
\def\CPC#1#2#3{{ Comput. Phys. Commun. }{\bf #1 }(#2) #3}
\def\etal{{\it et al.}}

\font\ninerm=cmr9

\begin{document}

\thispagestyle{empty}

\begin{titlepage}


\begin{center}
\vspace{2.1cm}
\boldmath
{\bf \Huge Search for charged Higgs bosons \\ 
in ${\mathrm e}^+{\mathrm e}^-$ collisions at energies\\ 
up to \sqs\ =  209~GeV}
\unboldmath
\vspace{1.8cm}

{\Large The ALEPH Collaboration$^*)$}

\end{center}

\vspace{1.5cm}
\begin{abstract}
\vspace{.5cm}
A search for charged Higgs bosons produced in pairs is performed
 with data collected at centre-of-mass energies
 ranging from 189 to 209\,GeV by ALEPH at LEP, corresponding to a total 
luminosity of 629\,\invpb.
The three final states \tntn, \tncs{} and \cscs{} are considered.
No evidence for a signal is found and
lower limits are set on the mass \mh{} as a
function of the branching fraction~\btn{}. 
In the framework of a two-Higgs-doublet model, and assuming \btn+\bcs=1,  
charged Higgs bosons with masses below 79.3\,\gevcc{} are 
excluded at $95\%$~confidence level independently of the branching ratios. 
\end{abstract}
\vfill
\vskip .2cm
\vskip .5cm
\noindent
--------------------------------------------\hfil\break
{\ninerm $^*)$ See next pages for the list of authors}
\end{titlepage}

\clearpage
\pagestyle{empty}
\newpage
\small
%
%
\newlength{\saveparskip}
\newlength{\savetextheight}
\newlength{\savetopmargin}
\newlength{\savetextwidth}
\newlength{\saveoddsidemargin}
\newlength{\savetopsep}
\setlength{\saveparskip}{\parskip}
\setlength{\savetextheight}{\textheight}
\setlength{\savetopmargin}{\topmargin}
\setlength{\savetextwidth}{\textwidth}
\setlength{\saveoddsidemargin}{\oddsidemargin}
\setlength{\savetopsep}{\topsep}
%
%
\setlength{\parskip}{0.0cm}
\setlength{\textheight}{25.0cm}
\setlength{\topmargin}{-1.5cm}
\setlength{\textwidth}{16 cm}
\setlength{\oddsidemargin}{-0.0cm}
\setlength{\topsep}{1mm}
\pretolerance=10000
\centerline{\large\bf The ALEPH Collaboration}
\footnotesize
\vspace{0.5cm}
{\raggedbottom
\begin{sloppypar}
\samepage\noindent
A.~Heister,
S.~Schael
\nopagebreak
\begin{center}
\parbox{15.5cm}{\sl\samepage
Physikalisches Institut das RWTH-Aachen, D-52056 Aachen, Germany}
\end{center}\end{sloppypar}
\vspace{2mm}
\begin{sloppypar}
\noindent
R.~Barate,
R.~Bruneli\`ere,
I.~De~Bonis,
D.~Decamp,
C.~Goy,
S.~Jezequel,
J.-P.~Lees,
F.~Martin,
E.~Merle,
\mbox{M.-N.~Minard},
B.~Pietrzyk,
B.~Trocm\'e
\nopagebreak
\begin{center}
\parbox{15.5cm}{\sl\samepage
Laboratoire de Physique des Particules (LAPP), IN$^{2}$P$^{3}$-CNRS,
F-74019 Annecy-le-Vieux Cedex, France}
\end{center}\end{sloppypar}
\vspace{2mm}
\begin{sloppypar}
\noindent
G.~Boix,$^{25}$
S.~Bravo,
M.P.~Casado,
M.~Chmeissani,
J.M.~Crespo,
E.~Fernandez,
M.~Fernandez-Bosman,
Ll.~Garrido,$^{15}$
E.~Graug\'{e}s,
J.~Lopez,
M.~Martinez,
G.~Merino,
A.~Pacheco,
D.~Paneque,
H.~Ruiz
\nopagebreak
\begin{center}
\parbox{15.5cm}{\sl\samepage
Institut de F\'{i}sica d'Altes Energies, Universitat Aut\`{o}noma
de Barcelona, E-08193 Bellaterra (Barcelona), Spain$^{7}$}
\end{center}\end{sloppypar}
\vspace{2mm}
\begin{sloppypar}
\noindent
A.~Colaleo,
D.~Creanza,
N.~De~Filippis,
M.~de~Palma,
G.~Iaselli,
G.~Maggi,
M.~Maggi,
S.~Nuzzo,
A.~Ranieri,
G.~Raso,$^{24}$
F.~Ruggieri,
G.~Selvaggi,
L.~Silvestris,
P.~Tempesta,
A.~Tricomi,$^{3}$
G.~Zito
\nopagebreak
\begin{center}
\parbox{15.5cm}{\sl\samepage
Dipartimento di Fisica, INFN Sezione di Bari, I-70126 Bari, Italy}
\end{center}\end{sloppypar}
\vspace{2mm}
\begin{sloppypar}
\noindent
X.~Huang,
J.~Lin,
Q. Ouyang,
T.~Wang,
Y.~Xie,
R.~Xu,
S.~Xue,
J.~Zhang,
L.~Zhang,
W.~Zhao
\nopagebreak
\begin{center}
\parbox{15.5cm}{\sl\samepage
Institute of High Energy Physics, Academia Sinica, Beijing, The People's
Republic of China$^{8}$}
\end{center}\end{sloppypar}
\vspace{2mm}
\begin{sloppypar}
\noindent
D.~Abbaneo,
P.~Azzurri,
T.~Barklow,$^{30}$
O.~Buchm\"uller,$^{30}$
M.~Cattaneo,
F.~Cerutti,
B.~Clerbaux,$^{23}$
H.~Drevermann,
R.W.~Forty,
M.~Frank,
F.~Gianotti,
T.C.~Greening,$^{26}$
J.B.~Hansen,
J.~Harvey,
D.E.~Hutchcroft,
P.~Janot,
B.~Jost,
M.~Kado,$^{2}$
P.~Mato,
A.~Moutoussi,
F.~Ranjard,
L.~Rolandi,
D.~Schlatter,
G.~Sguazzoni,
W.~Tejessy,
F.~Teubert,
A.~Valassi,
I.~Videau,
J.J.~Ward
\nopagebreak
\begin{center}
\parbox{15.5cm}{\sl\samepage
European Laboratory for Particle Physics (CERN), CH-1211 Geneva 23,
Switzerland}
\end{center}\end{sloppypar}
\vspace{2mm}
\begin{sloppypar}
\noindent
F.~Badaud,
S.~Dessagne,
A.~Falvard,$^{20}$
D.~Fayolle,
P.~Gay,
J.~Jousset,
B.~Michel,
S.~Monteil,
D.~Pallin,
J.M.~Pascolo,
P.~Perret
\nopagebreak
\begin{center}
\parbox{15.5cm}{\sl\samepage
Laboratoire de Physique Corpusculaire, Universit\'e Blaise Pascal,
IN$^{2}$P$^{3}$-CNRS, Clermont-Ferrand, F-63177 Aubi\`{e}re, France}
\end{center}\end{sloppypar}
\vspace{2mm}
\begin{sloppypar}
\noindent
J.D.~Hansen,
J.R.~Hansen,
P.H.~Hansen,
B.S.~Nilsson
\nopagebreak
\begin{center}
\parbox{15.5cm}{\sl\samepage
Niels Bohr Institute, 2100 Copenhagen, DK-Denmark$^{9}$}
\end{center}\end{sloppypar}
\vspace{2mm}
\begin{sloppypar}
\noindent
A.~Kyriakis,
C.~Markou,
E.~Simopoulou,
A.~Vayaki,
K.~Zachariadou
\nopagebreak
\begin{center}
\parbox{15.5cm}{\sl\samepage
Nuclear Research Center Demokritos (NRCD), GR-15310 Attiki, Greece}
\end{center}\end{sloppypar}
\vspace{2mm}
\begin{sloppypar}
\noindent
A.~Blondel,$^{12}$
\mbox{J.-C.~Brient},
F.~Machefert,
A.~Roug\'{e},
M.~Swynghedauw,
R.~Tanaka
\linebreak
H.~Videau
\nopagebreak
\begin{center}
\parbox{15.5cm}{\sl\samepage
Laoratoire Leprince-Ringuet, Ecole
Polytechnique, IN$^{2}$P$^{3}$-CNRS, \mbox{F-91128} Palaiseau Cedex, France}
\end{center}\end{sloppypar}
\vspace{2mm}
\begin{sloppypar}
\noindent
V.~Ciulli,
E.~Focardi,
G.~Parrini
\nopagebreak
\begin{center}
\parbox{15.5cm}{\sl\samepage
Dipartimento di Fisica, Universit\`a di Firenze, INFN Sezione di Firenze,
I-50125 Firenze, Italy}
\end{center}\end{sloppypar}
\vspace{2mm}
\begin{sloppypar}
\noindent
A.~Antonelli,
M.~Antonelli,
G.~Bencivenni,
F.~Bossi,
G.~Capon,
V.~Chiarella,
P.~Laurelli,
G.~Mannocchi,$^{5}$
G.P.~Murtas,
L.~Passalacqua
\nopagebreak
\begin{center}
\parbox{15.5cm}{\sl\samepage
Laboratori Nazionali dell'INFN (LNF-INFN), I-00044 Frascati, Italy}
\end{center}\end{sloppypar}
\vspace{2mm}
\begin{sloppypar}
\noindent
J.~Kennedy,
J.G.~Lynch,
P.~Negus,
V.~O'Shea,
A.S.~Thompson
\nopagebreak
\begin{center}
\parbox{15.5cm}{\sl\samepage
Department of Physics and Astronomy, University of Glasgow, Glasgow G12
8QQ,United Kingdom$^{10}$}
\end{center}\end{sloppypar}
\vspace{2mm}
\pagebreak
\begin{sloppypar}
\noindent
S.~Wasserbaech
\nopagebreak
\begin{center}
\parbox{15.5cm}{\sl\samepage
Department of Physics, Haverford College, Haverford, PA 19041-1392, U.S.A.}
\end{center}\end{sloppypar}
\vspace{2mm}
\begin{sloppypar}
\noindent
R.~Cavanaugh,$^{4}$
S.~Dhamotharan,$^{21}$
C.~Geweniger,
P.~Hanke,
V.~Hepp,
E.E.~Kluge,
G.~Leibenguth,
A.~Putzer,
H.~Stenzel,
K.~Tittel,
M.~Wunsch$^{19}$
\nopagebreak
\begin{center}
\parbox{15.5cm}{\sl\samepage
Kirchhoff-Institut f\"ur Physik, Universit\"at Heidelberg, D-69120
Heidelberg, Germany$^{16}$}
\end{center}\end{sloppypar}
\vspace{2mm}
\begin{sloppypar}
\noindent
R.~Beuselinck,
W.~Cameron,
G.~Davies,
P.J.~Dornan,
M.~Girone,$^{1}$
R.D.~Hill,
N.~Marinelli,
J.~Nowell,
S.A.~Rutherford,
J.K.~Sedgbeer,
J.C.~Thompson,$^{14}$
R.~White
\nopagebreak
\begin{center}
\parbox{15.5cm}{\sl\samepage
Department of Physics, Imperial College, London SW7 2BZ,
United Kingdom$^{10}$}
\end{center}\end{sloppypar}
\vspace{2mm}
\begin{sloppypar}
\noindent
V.M.~Ghete,
P.~Girtler,
E.~Kneringer,
D.~Kuhn,
G.~Rudolph
\nopagebreak
\begin{center}
\parbox{15.5cm}{\sl\samepage
Institut f\"ur Experimentalphysik, Universit\"at Innsbruck, A-6020
Innsbruck, Austria$^{18}$}
\end{center}\end{sloppypar}
\vspace{2mm}
\begin{sloppypar}
\noindent
E.~Bouhova-Thacker,
C.K.~Bowdery,
D.P.~Clarke,
G.~Ellis,
A.J.~Finch,
F.~Foster,
G.~Hughes,
R.W.L.~Jones,
M.R.~Pearson,
N.A.~Robertson,
M.~Smizanska
\nopagebreak
\begin{center}
\parbox{15.5cm}{\sl\samepage
Department of Physics, University of Lancaster, Lancaster LA1 4YB,
United Kingdom$^{10}$}
\end{center}\end{sloppypar}
\vspace{2mm}
\begin{sloppypar}
\noindent
O.~van~der~Aa,
C.~Delaere,
V.~Lemaitre
\nopagebreak
\begin{center}
\parbox{15.5cm}{\sl\samepage
Institut de Physique Nucl\'eaire, D\'epartement de Physique, Universit\'e Catholique de Louvain, 1348 Louvain-la-Neuve, Belgium}
\end{center}\end{sloppypar}
\vspace{2mm}
\begin{sloppypar}
\noindent
U.~Blumenschein,
F.~H\"olldorfer,
K.~Jakobs,
F.~Kayser,
K.~Kleinknecht,
A.-S.~M\"uller,
G.~Quast,$^{6}$
B.~Renk,
H.-G.~Sander,
S.~Schmeling,
H.~Wachsmuth,
C.~Zeitnitz,
T.~Ziegler
\nopagebreak
\begin{center}
\parbox{15.5cm}{\sl\samepage
Institut f\"ur Physik, Universit\"at Mainz, D-55099 Mainz, Germany$^{16}$}
\end{center}\end{sloppypar}
\vspace{2mm}
\begin{sloppypar}
\noindent
A.~Bonissent,
P.~Coyle,
C.~Curtil,
A.~Ealet,
D.~Fouchez,
P.~Payre,
A.~Tilquin
\nopagebreak
\begin{center}
\parbox{15.5cm}{\sl\samepage
Centre de Physique des Particules de Marseille, Univ M\'editerran\'ee,
IN$^{2}$P$^{3}$-CNRS, F-13288 Marseille, France}
\end{center}\end{sloppypar}
\vspace{2mm}
\begin{sloppypar}
\noindent
F.~Ragusa
\nopagebreak
\begin{center}
\parbox{15.5cm}{\sl\samepage
Dipartimento di Fisica, Universit\`a di Milano e INFN Sezione di
Milano, I-20133 Milano, Italy.}
\end{center}\end{sloppypar}
\vspace{2mm}
\begin{sloppypar}
\noindent
A.~David,
H.~Dietl,
G.~Ganis,$^{27}$
K.~H\"uttmann,
G.~L\"utjens,
W.~M\"anner,
\mbox{H.-G.~Moser},
R.~Settles,
G.~Wolf
\nopagebreak
\begin{center}
\parbox{15.5cm}{\sl\samepage
Max-Planck-Institut f\"ur Physik, Werner-Heisenberg-Institut,
D-80805 M\"unchen, Germany\footnotemark[16]}
\end{center}\end{sloppypar}
\vspace{2mm}
\begin{sloppypar}
\noindent
J.~Boucrot,
O.~Callot,
M.~Davier,
L.~Duflot,
\mbox{J.-F.~Grivaz},
Ph.~Heusse,
A.~Jacholkowska,$^{32}$
L.~Serin,
\mbox{J.-J.~Veillet},
J.-B.~de~Vivie~de~R\'egie,$^{28}$
C.~Yuan
\nopagebreak
\begin{center}
\parbox{15.5cm}{\sl\samepage
Laboratoire de l'Acc\'el\'erateur Lin\'eaire, Universit\'e de Paris-Sud,
IN$^{2}$P$^{3}$-CNRS, F-91898 Orsay Cedex, France}
\end{center}\end{sloppypar}
\vspace{2mm}
\begin{sloppypar}
\noindent
G.~Bagliesi,
T.~Boccali,
L.~Fo\`a,
A.~Giammanco,
A.~Giassi,
F.~Ligabue,
A.~Messineo,
F.~Palla,
G.~Sanguinetti,
A.~Sciab\`a,
R.~Tenchini,$^{1}$
A.~Venturi,$^{1}$
P.G.~Verdini
\samepage
\begin{center}
\parbox{15.5cm}{\sl\samepage
Dipartimento di Fisica dell'Universit\`a, INFN Sezione di Pisa,
e Scuola Normale Superiore, I-56010 Pisa, Italy}
\end{center}\end{sloppypar}
\vspace{2mm}
\begin{sloppypar}
\noindent
O.~Awunor,
G.A.~Blair,
G.~Cowan,
A.~Garcia-Bellido,
M.G.~Green,
L.T.~Jones,
T.~Medcalf,
A.~Misiejuk,
J.A.~Strong,
P.~Teixeira-Dias
\nopagebreak
\begin{center}
\parbox{15.5cm}{\sl\samepage
Department of Physics, Royal Holloway \& Bedford New College,
University of London, Egham, Surrey TW20 OEX, United Kingdom$^{10}$}
\end{center}\end{sloppypar}
\vspace{2mm}
\begin{sloppypar}
\noindent
R.W.~Clifft,
T.R.~Edgecock,
P.R.~Norton,
I.R.~Tomalin
\nopagebreak
\begin{center}
\parbox{15.5cm}{\sl\samepage
Particle Physics Dept., Rutherford Appleton Laboratory,
Chilton, Didcot, Oxon OX11 OQX, United Kingdom$^{10}$}
\end{center}\end{sloppypar}
\vspace{2mm}
\begin{sloppypar}
\noindent
\mbox{B.~Bloch-Devaux},
D.~Boumediene,
P.~Colas,
B.~Fabbro,
E.~Lan\c{c}on,
\mbox{M.-C.~Lemaire},
E.~Locci,
P.~Perez,
J.~Rander,
P. Seager,
B.~Tuchming,
B.~Vallage
\nopagebreak
\begin{center}
\parbox{15.5cm}{\sl\samepage
CEA, DAPNIA/Service de Physique des Particules,
CE-Saclay, F-91191 Gif-sur-Yvette Cedex, France$^{17}$}
\end{center}\end{sloppypar}
\vspace{2mm}
\begin{sloppypar}
\noindent
N.~Konstantinidis,
A.M.~Litke,
G.~Taylor
\nopagebreak
\begin{center}
\parbox{15.5cm}{\sl\samepage
Institute for Particle Physics, University of California at
Santa Cruz, Santa Cruz, CA 95064, USA$^{22}$}
\end{center}\end{sloppypar}
\vspace{2mm}
\begin{sloppypar}
\noindent
C.N.~Booth,
S.~Cartwright,
F.~Combley,$^{31}$
P.N.~Hodgson,
M.~Lehto,
L.F.~Thompson
\nopagebreak
\begin{center}
\parbox{15.5cm}{\sl\samepage
Department of Physics, University of Sheffield, Sheffield S3 7RH,
United Kingdom$^{10}$}
\end{center}\end{sloppypar}
\vspace{2mm}
\begin{sloppypar}
\noindent
A.~B\"ohrer,
S.~Brandt,
C.~Grupen,
J.~Hess,
A.~Ngac,
G.~Prange,
U.~Sieler
\nopagebreak
\begin{center}
\parbox{15.5cm}{\sl\samepage
Fachbereich Physik, Universit\"at Siegen, D-57068 Siegen, Germany$^{16}$}
\end{center}\end{sloppypar}
\vspace{2mm}
\begin{sloppypar}
\noindent
C.~Borean,
G.~Giannini
\nopagebreak
\begin{center}
\parbox{15.5cm}{\sl\samepage
Dipartimento di Fisica, Universit\`a di Trieste e INFN Sezione di Trieste,
I-34127 Trieste, Italy}
\end{center}\end{sloppypar}
\vspace{2mm}
\begin{sloppypar}
\noindent
H.~He,
J.~Putz,
J.~Rothberg
\nopagebreak
\begin{center}
\parbox{15.5cm}{\sl\samepage
Experimental Elementary Particle Physics, University of Washington, Seattle,
WA 98195 U.S.A.}
\end{center}\end{sloppypar}
\vspace{2mm}
\begin{sloppypar}
\noindent
S.R.~Armstrong,
K.~Berkelman,
K.~Cranmer,
D.P.S.~Ferguson,
Y.~Gao,$^{29}$
S.~Gonz\'{a}lez,
O.J.~Hayes,
H.~Hu,
S.~Jin,
J.~Kile,
P.A.~McNamara III,
J.~Nielsen,
Y.B.~Pan,
\mbox{J.H.~von~Wimmersperg-Toeller}, 
W.~Wiedenmann,
J.~Wu,
Sau~Lan~Wu,
X.~Wu,
G.~Zobernig
\nopagebreak
\begin{center}
\parbox{15.5cm}{\sl\samepage
Department of Physics, University of Wisconsin, Madison, WI 53706,
USA$^{11}$}
\end{center}\end{sloppypar}
\vspace{2mm}
\begin{sloppypar}
\noindent
G.~Dissertori
\nopagebreak
\begin{center}
\parbox{15.5cm}{\sl\samepage
Institute for Particle Physics, ETH H\"onggerberg, 8093 Z\"urich,
Switzerland.}
\end{center}\end{sloppypar}
}
\footnotetext[1]{Also at CERN, 1211 Geneva 23, Switzerland.}
\footnotetext[2]{Now at Fermilab, PO Box 500, MS 352, Batavia, IL 60510, USA}
\footnotetext[3]{Also at Dipartimento di Fisica di Catania and INFN Sezione di
 Catania, 95129 Catania, Italy.}
\footnotetext[4]{Now at University of Florida, Department of Physics, Gainesville, Florida 32611-8440, USA}
\footnotetext[5]{Also Istituto di Cosmo-Geofisica del C.N.R., Torino,
Italy.}
\footnotetext[6]{Now at Institut f\"ur Experimentelle Kernphysik, Universit\"at Karlsruhe, 76128 Karlsruhe, Germany.}
\footnotetext[7]{Supported by CICYT, Spain.}
\footnotetext[8]{Supported by the National Science Foundation of China.}
\footnotetext[9]{Supported by the Danish Natural Science Research Council.}
\footnotetext[10]{Supported by the UK Particle Physics and Astronomy Research
Council.}
\footnotetext[11]{Supported by the US Department of Energy, grant
DE-FG0295-ER40896.}
\footnotetext[12]{Now at Departement de Physique Corpusculaire, Universit\'e de
Gen\`eve, 1211 Gen\`eve 4, Switzerland.}
\footnotetext[13]{Supported by the Commission of the European Communities,
contract ERBFMBICT982874.}
\footnotetext[14]{Supported by the Leverhulme Trust.}
\footnotetext[15]{Permanent address: Universitat de Barcelona, 08208 Barcelona,
Spain.}
\footnotetext[16]{Supported by Bundesministerium f\"ur Bildung
und Forschung, Germany.}
\footnotetext[17]{Supported by the Direction des Sciences de la
Mati\`ere, C.E.A.}
\footnotetext[18]{Supported by the Austrian Ministry for Science and Transport.}
\footnotetext[19]{Now at SAP AG, 69185 Walldorf, Germany}
\footnotetext[20]{Now at Groupe d' Astroparticules de Montpellier, Universit\'e de Montpellier II, 34095 Montpellier, France.}
\footnotetext[21]{Now at BNP Paribas, 60325 Frankfurt am Mainz, Germany}
\footnotetext[22]{Supported by the US Department of Energy,
grant DE-FG03-92ER40689.}
\footnotetext[23]{Now at Institut Inter-universitaire des hautes Energies (IIHE), CP 230, Universit\'{e} Libre de Bruxelles, 1050 Bruxelles, Belgique}
\footnotetext[24]{Also at Dipartimento di Fisica e Tecnologie Relative, Universit\`a di Palermo, Palermo, Italy.}
\footnotetext[25]{Now at McKinsey and Compagny, Avenue Louis Casal 18, 1203 Geneva, Switzerland.}
\footnotetext[26]{Now at Honeywell, Phoenix AZ, U.S.A.}
\footnotetext[27]{Now at INFN Sezione di Roma II, Dipartimento di Fisica, Universit\`a di Roma Tor Vergata, 00133 Roma, Italy.}
\footnotetext[28]{Now at Centre de Physique des Particules de Marseille, Univ M\'editerran\'ee, F-13288 Marseille, France.}
\footnotetext[29]{Also at Department of Physics, Tsinghua University, Beijing, The People's Republic of China.}
\footnotetext[30]{Now at SLAC, Stanford, CA 94309, U.S.A.}
\footnotetext[31]{Deceased.}
\footnotetext[32]{Also at Groupe d' Astroparticules de Montpellier, Universit\'e de Montpellier II, 34095 Montpellier, France.}  
\setlength{\parskip}{\saveparskip}
\setlength{\textheight}{\savetextheight}
\setlength{\topmargin}{\savetopmargin}
\setlength{\textwidth}{\savetextwidth}
\setlength{\oddsidemargin}{\saveoddsidemargin}
\setlength{\topsep}{\savetopsep}
\normalsize
\newpage
\pagestyle{plain}
\setcounter{page}{1}

\pagestyle{plain}
\pagenumbering{arabic}

\section{Introduction}
 The Standard Model of electroweak interactions requires
only one doublet of complex scalar fields, resulting in
a single neutral Higgs particle. 
The simplest extensions of the Standard Model assume
two complex scalar-field doublets, with a total of eight degrees
of freedom. As in the Standard Model, three of the degrees of freedom
are associated with the longitudinal components of the 
$\mathrm{W}^{\pm}$ and Z bosons.
The remaining five degrees of freedom appear as five
physical scalar Higgs states: three neutral Higgs bosons
and the charged Higgs bosons $\mathrm{H}^{\pm}$.


In the two-Higgs-doublet case, 
the charged Higgs boson couplings are completely specified in terms of
the electric charge and the weak mixing angle
$\mathrm{\theta_W}$. The production cross-section thus
depends only on the mass \mh{}. For masses 
accessible at LEP\,2 energies, the charged Higgs boson decays 
with negligible lifetime and width into either 
$\mathrm{c\bar{s}}$/$\mathrm{c\bar{b}}$ or 
${\mathrm \tau^+\nu_\tau}$ final states. Because the analyses are not 
sensitive to the quark flavour, and because the $\mathrm{c\bar{s}}$ 
decay mode dominates over $\mathrm{c\bar{b}}$, 
$\mathrm{c\bar{s}}$ stands for either $\mathrm{c\bar{s}}$ or
$\mathrm{c\bar{b}}$ in the following.
Therefore, ~\btn+\bcs=1 is 
assumed and $\hh$ pair production leads to three 
final states (\tntn{}, \tncs{}/\cstn{} and \cscs) for which separate searches
are performed. 

  The ALEPH data collected at energies up to 189 GeV have already
been analysed and the search results published
in Refs.~\cite{HCH172,HCH183,HCH189}.  The negative result of
the search, under the hypotheses specified above,
was translated into a  lower  limit on the $\hpm$ mass
of 65.5~GeV/c$^2$ at 95\% confidence level (C.L.). 
Results from other experiments 
are given in Ref.~\cite{ADL189}. The present letter describes 
the search for pair-produced charged Higgs bosons using the data collected 
up to the end of data taking. 
An improved analysis has been designed for the fully leptonic channel.
In the semileptonic search,
the rejection of the $\mathrm{W^+W^-}$ background has been refined
with a method based on
a combination of the charge-tagged boson
production angle and a $\tau$ polarization estimator.
For the four-jet event selection, the linear discriminant analysis (LDA)
has been re-optimized to account for the additional integrated luminosity 
collected at increased centre-of-mass energies.


\section{The ALEPH detector and event samples}

A complete and detailed description of the ALEPH detector and its
performance, as well as of the standard reconstruction and analysis
algorithms can be found in Refs.~\cite{ALEDEC,ALEPER}. Only those items
relevant for the final states under study in this letter are summarized
below.

The trajectories of the charged particles (called {\it charged tracks} in
the following) are measured with the central tracking system, formed by
a silicon vertex detector, an inner drift chamber and a large time
projection chamber, all immersed in the 1.5\,T axial magnetic field from
a superconducting solenoidal coil. Electrons and photons are identified
in the electromagnetic calorimeter, a highly segmented sampling calorimeter
placed between the tracking device and the coil. Muons are identified
in the hadron calorimeter, a 1.2\,m thick iron yoke instrumented
with 23 layers of streamer tubes, surrounded with two double layers of muon
chambers. Together with the luminometers, the hermetic calorimetric coverage
extends down to 34\,mrad of the beam axis. The missing energy and momentum
from, {\it e.g.}, tau charged Higgs boson decays, are determined
with an energy-flow algorithm which combines particle identification,
tracking and calorimetry information into a set of energy-flow particles,
used in the present analyses.

The data analysed in this letter were collected at LEP between 1998 and 2000
at ${\rm e}^+{\rm e}^-$ centre-of-mass energies ranging from
189~to~209\,GeV,
corresponding to a total integrated luminosity of 629\,${\rm pb}^{-1}$. The
details for each sample 
are given in Table~\ref{tab:lumi}.

\begin{table}[htbp]
\begin{center}
\caption{\footnotesize Integrated luminosities, centre-of-mass energy ranges
and mean centre-of-mass energy values for the data collected with the ALEPH
detector from 1998  to 2000.}
\begin{tabular}{|c|c|c|c|}
  \multicolumn{4}{c}{} \\  \hline \hline
 Year & Luminosity (${\rm pb}^{-1}$) & Energy range (GeV) & $\langle
\sqrt{s} \rangle$ (GeV) \\ \hline
 2000 &   217.2            & $204-209$        & 206.1 \\  \hline
 1999 &   42.0            &   $-$              & 201.6 \\
      &   86.3            &   $-$              & 199.5 \\
      &   79.8            &   $-$              & 195.5 \\
      &   28.9            &   $-$              & 191.6 \\ \hline
 1998 &  174.4            &   $-$              & 188.6 \\
\hline
\hline
\end{tabular}
\label{tab:lumi}
\end{center}
\end{table}

Fully simulated samples of events reconstructed with the same programs 
as the data
were used for the background estimates, the design of the selections and
the optimization of the selection cuts. The most important background
sources are {\it (i)} difermion events (${\rm e}^+{\rm e}^- \to
\tau^+\tau^-$
and ${\rm q\bar q}$) simulated with the {\tt KORALZ}~\cite{WAS1}
generator;
and {\it (ii)} ${\rm e}^+{\rm e}^- \to {\rm W}^+{\rm W}^-$ and other
four-fermion processes simulated with the {\tt KORALW}~\cite{KORALW}
and {\tt PYTHIA}~\cite{PYTHIA} generators. Event samples of these background
processes, corresponding to at least 20 times the collected luminosity,
were generated. The ${\rm W}^+{\rm W}^-$ cross sections predicted by
{\tt RACOONWW}~\cite{racoon} and {\tt YFSWW}~\cite{yfsww} were used
as discussed in Ref.~\cite{red2.3}. Finally, the two-photon interactions
($\gamma\gamma \to$~leptons) were simulated with the {\tt
PHOT02}~\cite{PHOT02}
generator. Samples of these events with at least six times the collected
luminosity were generated.

The signal events generated with the {\tt HZHA}~\cite{JANOT} program
were simulated for each of the final states and centre-of-mass energies
(Table~\ref{tab:lumi}), and for charged Higgs boson masses between
45~and~100\,GeV/$c^2$.

\section{Analyses}
        An event selection has been defined for each of the \tntn, 
\tncs{}/\cstn{} (hereafter referred to as \tncs{}) and \cscs{} channels, and
was optimized for \btn{} = $100\%$, $50\%$
and $0\%$, respectively.
The selection criteria
were chosen to achieve the highest $95\%$ confidence level 
expected limit on the charged 
Higgs boson mass in the absence of signal.

\begin{boldmath}
\subsection{The \tntn{} final state}
\end{boldmath}
Events with two to six charged
tracks (at least one and at most four of each sign) are considered. 
Leptonic events 
$\ww \to \ell \nu \ell^\prime \bar{\nu}$ ($\ell, \ell^\prime = 
\mathrm{e}$ or $\mu$)
are rejected by requiring that the momentum of any identified electron or 
muon be less than $0.1 \sqrt{s}$. The events are then forced to form
two jets with the JADE algorithm~\cite{JADE}.
An event is selected if both jet polar angles $\theta_{1,2}$
satisfy $| \cos \theta _{1,2} | < 0.96$, if their reconstructed
masses are less than 3\,\gevcc~and if each jet contains at least one 
charged track. 
To suppress the high cross section $\gamma \gamma \to \ff$
processes, the total visible mass is required to be in excess 
of $0.075 \sqrt{s}$, the momentum
transverse to the beam is required to be greater than 10\,\gevc, and there
must be no energy deposited in a cone of $12^{\circ}$ around the beam axis.
The signal selection efficiency of the
latter cut is corrected for the effect of the beam-related
background, not included
in the simulation, and is estimated from events triggered at random
beam crossings.
The  relative loss of signal efficiency is about 7\%.

\begin{figure}[h]
\centering
  \epsfysize=9.5cm
  \leavevmode
  \epsfbox[-50 410 596 670]{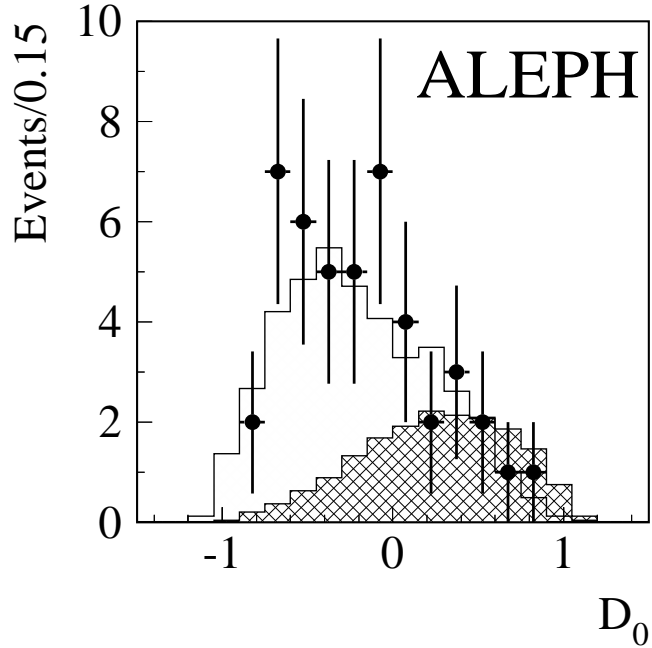}
 \caption{\small{The distribution of the discriminant variable $D_0$ described
 in the text for the fully-leptonic channel. The points are the data, the
 open histogram is the Standard Model background and the hatched histogram 
 represents the Higgs signal expectation, absolutely normalized, 
with $\mh$~=~85\,\gevcc{}. }}
 \label{fig:hphm_tntn1}
 \end{figure}

Nearly coplanar
tau pairs from $\eett{} (\gamma)$ are rejected by
requiring that the angle $\alpha$ between the two tau jets be 
less than $170^{\circ}$ and the angle between the projections 
of their momenta onto the plane
transverse to the beam axis be less than $165^{\circ}$. The missing energy is 
required to be greater than 80\,\gev{} and the missing mass greater than 
70\,\gevcc{}. In order to
improve the $\ww$ background rejection, an LDA has been used to construct a
discriminant variable $D_0$ from a combination of the following four 
quantities:
\begin{itemize}
\item a charge-tagged angular variable calculated from
the polar angles of the $\tau$ jets   
and their charges as
 $C \; = \; \frac{1}{2} \, \left[ \, Q_{1} \cos\theta_1 \; + \;
 Q_{2} \cos\theta_2 \, \right];$
\item the angle $\alpha$ between the two tau jets;
\item the missing transverse momentum of the event $P_T^{\mathrm{miss}};$
\item the value $y_{23}$ of the jet-clustering resolution parameter
for which the transition from two to three jets occurs.
\end{itemize}
The optimal discriminant variable was found to be  
\begin{eqnarray}
D_0 & = & 0.930      \,\, C \;
         - \; 0.250 \,\, \alpha  \;
         + \; 0.008   \,\,  P_T ^{\mathrm{miss}} \;
         - \; 110   \,\,  y_{23} \;
         + \; 0.426\; ,
\nonumber  
\end{eqnarray}
where $\alpha$ is in radians
and $P_T ^{\mathrm{miss}}$ in \gevc{}.
The distribution of $D_0$ is displayed in Fig.~\ref{fig:hphm_tntn1}. This 
quantity is used as a discriminant variable in the derivation of the mass
limit.

The signal event selection efficiencies, parametrized as a function of $\mh$,
are given in Table~\ref{eff} for 
$\sqrt{s}=206$\,GeV. 
The selection efficiencies are
almost independent of the centre-of-mass energy and increase 
only slightly with $\mh$. 
For a signal with $\mh$=85\,\gevcc{} and \btn=1, 
a total of 16.5 events is expected in
the data taken at centre-of-mass energies between 189\,GeV and 209\,GeV.
\begin{table}[t]
 \caption{\small{The signal event selection efficiencies $\epsilon$ (in \%), 
parametrized as a function 
of the charged Higgs boson mass $\mh$, at \sqs{}= 206\,GeV.}}
\begin{center}
 \begin{tabular}{|l|c|c|c|c|c|c|c|}
\hline
$\mh$(\gevcc{}) & 60 & 65 & 70 & 75 & 80 & 85 & 90 \\
\hline
\hline
$\epsilon \ (\tntn)$ & 24.4 & 25.5 & 26.4 & 27.3 & 28.0 & 28.5 & 28.9 \\
\hline
$\epsilon \ (\tncs)$ & 49.1 & 48.0 & 45.8 & 42.8 & 38.8 & 33.9 & 28.0 \\
\hline
$\epsilon \ (\cscs)$ & 60.7 & 62.9 & 64.5 & 65.5 & 66.1 & 66.3 & 66.3 \\
\hline
\end{tabular}
\label{eff}
\end{center}
\end{table} 
The numbers of events selected are given in Table~\ref{yields}, 
compared to the 
expectations from the Standard Model backgrounds,
dominated by \ww{} production.  

\begin{table}[t]
 \caption{\small{Numbers  of candidate events and background expected
from Standard Model processes, for each of the three years of data taking.}}
\begin{center}
 \begin{tabular}{|l|c|c|c|}
\hline
 Channel   & \sqs{}    & observed   & expected    \\
           &  (GeV)    & events     & background  \\ 
\hline
\hline
$ \tntn$   &  188.6    &  14       &  11.0      \\
           & 192-202   &  22       &  15.6      \\
           & 204-209   &   9       &  14.0      \\
\hline
$ \tncs$   & 188.6     &  63       &  67.3      \\
           & 192-202   &  89       & 113.1      \\
           & 204-209   & 127       & 108.9      \\ 
\hline
$ \cscs$   & 188.6     &  778      & 826.3      \\
           & 192-202   &  1034     & 1102.6     \\
           & 204-209   &  950      & 963.2      \\ 
\hline
\end{tabular}
\label{yields}
\end{center}
\end{table}

The systematic uncertainty on the number of expected signal events is
 estimated 
to be 3.1\%, dominated by the effect of limited  Monte Carlo statistics (2.4\%)
and the uncertainty on the cross section for charged Higgs boson 
production (2\%). The systematic error on the background level is 
estimated to be 1.5\%, dominated by the effects of limited Monte Carlo 
statistics (1.3\%), by the uncertainty on the cross section 
for the $\ww$ process (0.5\%) and the 
uncertainty on the cross section for two-photon production (5\%). 


\begin{boldmath}
\subsection{The \tncs{} final state}
\end{boldmath}

The mixed final state \tncs{} is characterized by two jets originating 
from the hadronic decay of one of the charged Higgs bosons
and a $\tau$ jet with missing energy due to the prompt neutrino as well 
as to the neutrino(s) from the subsequent  $\tau$ decay. 

The preselection is the same as that described in Ref.~\cite{HCH189}. In 
order to identify the $\tau$ jet an algorithm based on ``minijets'' is 
used as described in Ref.~\cite{minijets}. 
%
%
If a minijet satisfies the $\tau$-jet selection criteria, the rest 
of the event is clustered into two jets using the Durham~\cite{durham} 
clustering algorithm.
A kinematic fit is 
performed with the constraints of energy and momentum conservation and 
equality of the \cs{} and \tn{} masses. If there is more than 
one $\tau$ candidate the combination with the lowest 
$\chi^2$ is taken.

In order to reject background from \wwemnqq{}, 
the measured energy of the $\tau$ jet boosted into the Higgs rest 
frame is required to be less than $0.175 \sqrt{s}$. The boost is 
performed using the information from the hadronic side of the event.

After this procedure the following four variables are chosen to further
suppress the background:

\begin{itemize}
\item the total missing transverse momentum of the event, 
$P_{T}^{\mathrm{miss}} ;$
\item the isolation angle $\theta_{\mathrm{iso}}$ of the $\tau$, 
defined as the half-angle 
of the cone around the $\tau$ jet direction containing 5\% of the total 
energy of the rest of the event;
\item the $\chi^2$ from the kinematic fit;
\item the decay angle $\theta^{\mathrm{ch}}_{\tau}$, defined as
the angle between the $\tau$ momentum 
in the Higgs boson centre-of-mass frame 
and the Higgs boson flight direction, 
charge-tagged with the charge of the $\tau$, to exploit the 
asymmetry in the W system, absent for scalars.
\end{itemize}
The four variables are linearly combined into one variable, $D_1$, defined as
\begin{eqnarray}
D_1 & = & 0.021 \,\, P_{T}^{\mathrm{miss}} \; + \; 0.400 \,\, 
\theta_{\mathrm{iso}} \; - \;
  0.058 \,\, \chi^2\; - \; 0.148 \, \, 
\theta^{\mathrm{\mathrm{ch}}}_{\tau} \; - \;
  0.881\;  \nonumber 
\end{eqnarray}
where $P_{T}^{\mathrm{miss}}$  is in \gevc, and $\theta_{\mathrm{iso}}$
and $\theta^{\mathrm{ch}}_{\tau}$ are in radians.
Events are selected by requiring that $D_1>-0.1$. The background 
consists primarily of \wwlnqq{} events.

\begin{figure}
\centering
  \epsfysize=8.0cm
  \leavevmode
  \epsfbox[86 410 546 670]{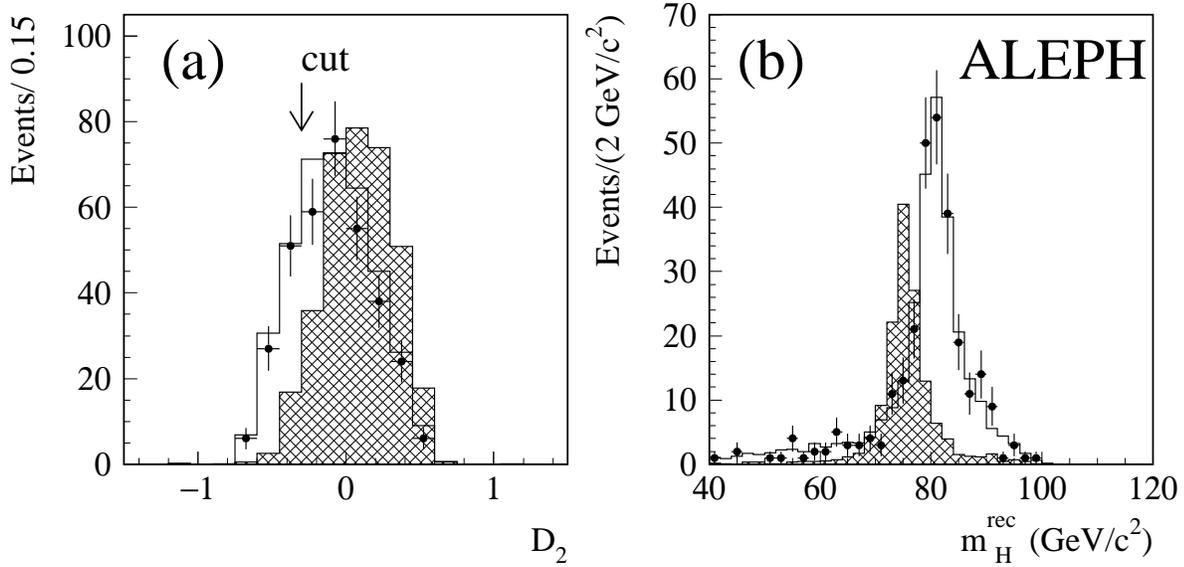}
 \caption{(a) {\small The distribution of the discriminant variable $D_2$ 
described in the text for the semi-leptonic channel.} 
(b) {\small The distribution of the fitted mass of the 
Higgs boson candidates after the cut on $D_2$.
The points are the data, the 
open histogram is the Standard Model background and the hatched histogram 
represents the Higgs boson signal expectation with $\mh$~=~75\,\gevcc{}. 
The signal is arbitrarily normalized.}}
 \label{fig:hphm_tncs1}
\end{figure}

Due to the scalar nature of the $\hp$,
the $\tau^+$ from its decay is produced in a left-handed 
helicity state, in contrast to the
$\tau^+$'s from $\mathrm{W}^+$ decays.
Variables designed for the measurement of the 
$\tau$ polarization at LEP\,1~\cite{taupol} have been used to form an 
event-by-event helicity estimator, ${\cal E}_{\tau}$.
This
variable, together with the charge-tagged production angle
$\theta_{\mathrm{prod}}^{\mathrm{ch}}$~\cite{HCH189}, is used to 
discriminate further between \wwtnqq{} and $\hh\to\tncs{}$ events. The two 
variables are combined into another variable, $D_2$, defined as
\begin{eqnarray}
D_2 & = & -\; 0.461 \,\, \theta_{\mathrm{prod}}^{\mathrm{ch}} \; - \;
        0.517 \,\, {\cal E}_{\tau} \; + \; 1.020\; , \nonumber
\end{eqnarray}
where $\theta_{\mathrm{prod}}^{\mathrm{ch}}$ is expressed in radians.
The distribution of $D_2$ is shown in Fig.~\ref{fig:hphm_tncs1}a.
The cut optimization yields $D_2>-0.3$ for $\mh$=75\,\gevcc.
The selection efficiencies are given
in Table~\ref{eff} as a function of the Higgs boson mass for 
$\sqrt{s}=206$\,GeV. 
They are only weakly dependent on $\sqrt{s}$.
In the data collected between $\sqrt{s}=189$ and 209\,GeV,
the numbers of selected events are compared with the 
background expectations in Table~\ref{yields}.
The fitted-mass distribution of the Higgs boson
candidates is shown in Fig.~\ref{fig:hphm_tncs1}b. 
For \mh=77\,\gevcc{}, close to the sensitivity of this search,
and for \btn=0.5, a total of 
21.2 signal events is expected.


The systematic uncertainty on the number of expected signal events is 
estimated to be 3.0\%. The main contributions are the finite size of the 
simulated event samples (2.2\%), calorimeter
calibration uncertainties (0.5\%) and the uncertainty on the
cross section for charged Higgs boson
production (2\%). The systematic error on the  background level was 
estimated to be 3.9\%. The main contributions are from limited
statistics of the simulated event samples (2.5\%), 
uncertainty on the cross section for the $\ww$ process 
(0.5\%) and calibration uncertainties (3\%). 

\begin{boldmath}
\subsection{The \cscs{} final state}
\end{boldmath}
 The hadronic decays of pair-produced charged Higgs bosons lead 
to a four-jet final state with equal mass
dijet systems. The preselection remains unchanged with
respect to Ref.~\cite{HCH189}. 

 A five-constraint kinematic fit is performed with energy-momentum
conservation and equal dijet-mass constraints. 
In this fitting procedure, the errors on the jet energies and angles are 
parametrized as for the W mass measurement in the four-jet 
channel~\cite{WMASS}.
The pairing is chosen as the dijet combination giving  the minimum~$\chi^2$.

To evaluate the mass difference between the two dijet invariant masses,
momentum and energy conservation is imposed to rescale
the energies of the four jets, fixing the jet velocities at their measured 
values. The  mass difference $\Delta m$ between the two rescaled dijets is
required to be  smaller than 30\,\gevcc{}.

 To improve the background rejection a linear discriminant
$D_3$  is constructed, combining the following five variables:

\begin{itemize}
\item the production polar angle $\theta_{\mathrm{prod}}$, 
{\it i.e.} the angle between the Higgs boson 
momentum direction and the beam axis;
\item the difference $\Delta m$ between the two rescaled dijet masses;
\item the $\chi^2$ of the 5C kinematic fit; 
\item the product of the minimum jet energy $E_{\mathrm{min}}$ and the minimum
 jet-jet angle $\theta_\qq{}$;
\item the logarithm of the
QCD four-jet matrix element squared ${\cal M}_{\mathrm{QCD}}$~\cite{matele}.
\end{itemize}

\begin{figure}
\centering
   \epsfysize=8cm
   \leavevmode
\epsfbox[86 410 546 670]{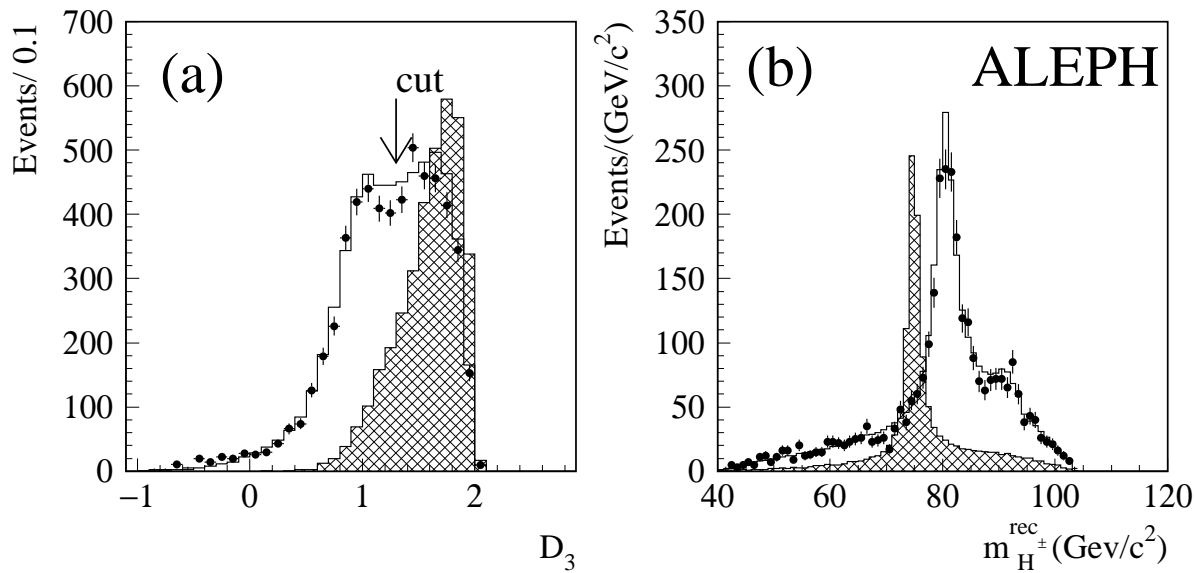}
\caption{(a) {\small The distribution of the discriminating variable $D_3$.}
(b) 
{\small The distribution of the reconstructed mass of the Higgs boson
candidates after the cut on the discriminating variable. 
The points are the data, the open histograms  are the Standard Model
backgrounds and the hatched histogram represents the  Higgs signal expectation
for
$\mh$~=~75\,\gevcc{}. 
The signal is arbitrarily normalized.}}
\label{fig:hphm_cscs1}
\end{figure}

 The optimized LDA coefficients were determined at
\sqs{}~= 206~GeV with a cocktail of five charged Higgs boson masses ranging
between 80 and 88~\gevcc{}, leading to:
\begin{eqnarray}     
D_3 & = & - \; 0.951  \, \, \cos^2 \theta_{\mathrm{prod}} \;  
        - \; 0.0065 \, \, \Delta m \; 
        - \; 0.000968 \, \, \chi^2_{\mathrm{5C}}  \nonumber \\
  &   & \; - \; 0.0034 \, \, (E_{\mathrm{min}} \times \theta_\qq) \;
        - \; 0.335  \, \, \log_{10}({\cal M}_{\mathrm{QCD}})\; , \nonumber  
\end{eqnarray}
with $\Delta m$ in \gevcc{}, $E_{\mathrm{min}}$ in GeV, $\theta_\qq$ in
radians, and ${\cal M}_{\mathrm{QCD}}$ in $\mathrm{GeV}^{-4}$.
The distribution of $D_3$ is shown in Fig.~\ref{fig:hphm_cscs1}a. The cut 
was optimized for $\mh$=76, 80 and 84\,\gevcc{}. Events are 
accepted if $D_3>1.3$. For $\mh$=75\,\gevcc{} and \btn=0, a total of
101.9 events is expected in the data.
The efficiency does not depend on $\sqrt{s}$.

 After the complete selection, the comparison between data and simulation is 
displayed in Fig.~\ref{fig:hphm_cscs1}b for the dijet invariant mass. 
The numbers of events observed in the data are compared in Table~\ref{yields} 
to the expected background  from Standard Model processes, dominated by
$\ww$ production.
An overall 2.4 standard deviation deficit with respect to 
expectation is observed. 
It is correlated with the deficit observed in the measurement of the $\ww$
hadronic cross section~\cite{red2.3}, which was ascribed to a
statistical fluctuation. 

The systematic error on the number of expected signal events is estimated 
to be 2.5\%.  The main contributions are from limited sample 
statistics (1.3\%),  uncertainty on the 
cross section for charged Higgs production (2\%) and
accuracy of the simulation (0.5\%).
The systematic error on the expected background, dominated by \ww{}
and \qq{} production, is estimated to be 2.0\%.
The main contributions are from the simulated sample statistics 
(0.4\% for \ww{} and 1.6\% for \qq{}), the uncertainty on the cross section 
(0.5\% for \ww{} and 5\% for \qq{}), and the adequacy of 
the simulation (1.4\% for \ww{} and 2.1\% for \qq{}). 

\section{\label{combine}Results}

No evidence for a signal is observed in the data. The results of the three 
selections have been combined  to set a 
95\% C.L. lower limit on the mass of charged Higgs bosons. 

Full background subtraction has been performed in setting the limit with 
the likelihood ratio test statistic~\cite{Eadie}. 
Systematic uncertainties are taken into account 
according to Ref.~\cite{C+H}.
To improve the sensitivity of the analysis, the
charged Higgs boson mass has been used as a 
discriminating variable for the \cscs{} and \tncs{} channels. In the 
previous publications~\cite{HCH172,HCH183,HCH189}, only event counting was 
used in the 
\tntn{} channel. In this analysis, the discriminant variable $D_0$ has 
been introduced in the limit setting procedure.

\begin{figure}[ht]
\begin{center}
  \epsfxsize=12.0cm
  \leavevmode
  \epsfbox[24 147 538 668]{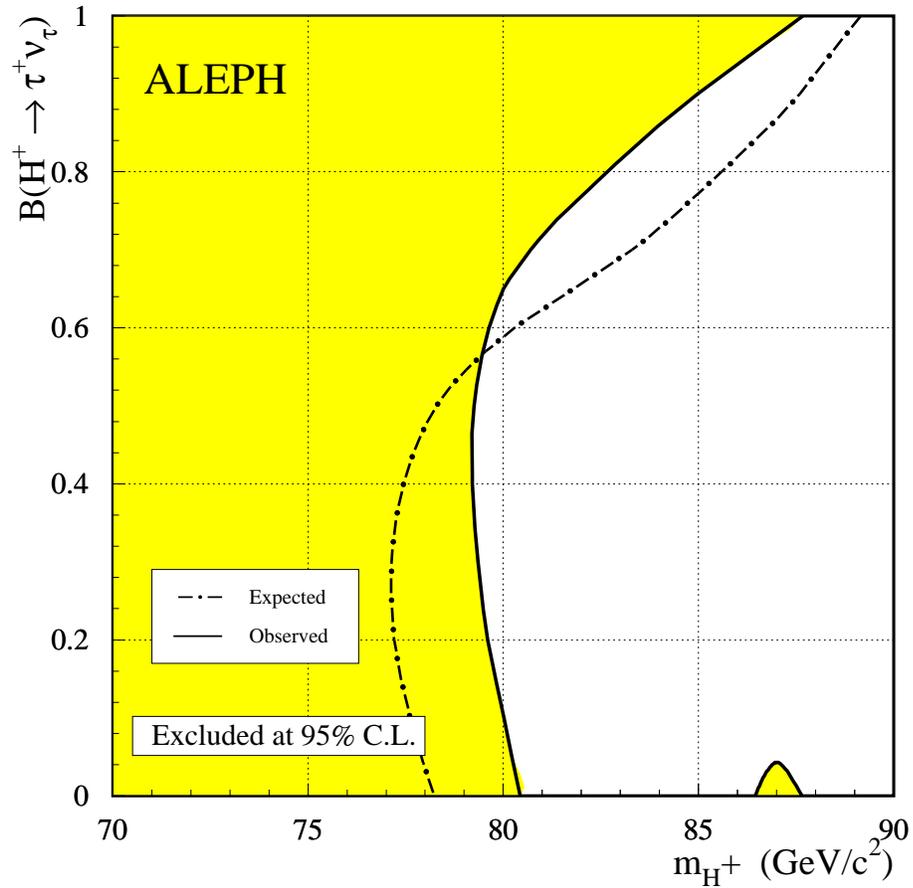}
\caption{\small Limit at $95\%$ C.L. on the charged
Higgs boson mass as a function
of \btn. The expected (dash-dotted) 
and observed (solid) exclusion curves are shown 
for the combination of the three  
analyses, using the full 189--209\,GeV data set.}
\label{combi}
\end{center}
\end{figure}

The result of the combination of the three analyses is shown in 
Fig.~\ref{combi}.
Charged Higgs bosons with mass lower than 79.3\,\gevcc{} are excluded
at the 95\% C.L. independently of \btn. 
The corresponding expected exclusion is 77.1\,\gevcc{}.
For the values $\btn=0$  and 1, 95\% C.L. lower limits on $\mh$ are set
at 80.4\,$\gevcc$ (with 78.2\,$\gevcc$ expected) 
and 87.8\,$\gevcc$ (with 89.2\,$\gevcc$ expected)
respectively.

\begin{figure}[h]
\centering
  \epsfxsize=0.8\textwidth
  \leavevmode
  \epsfbox[18 142 533 677]{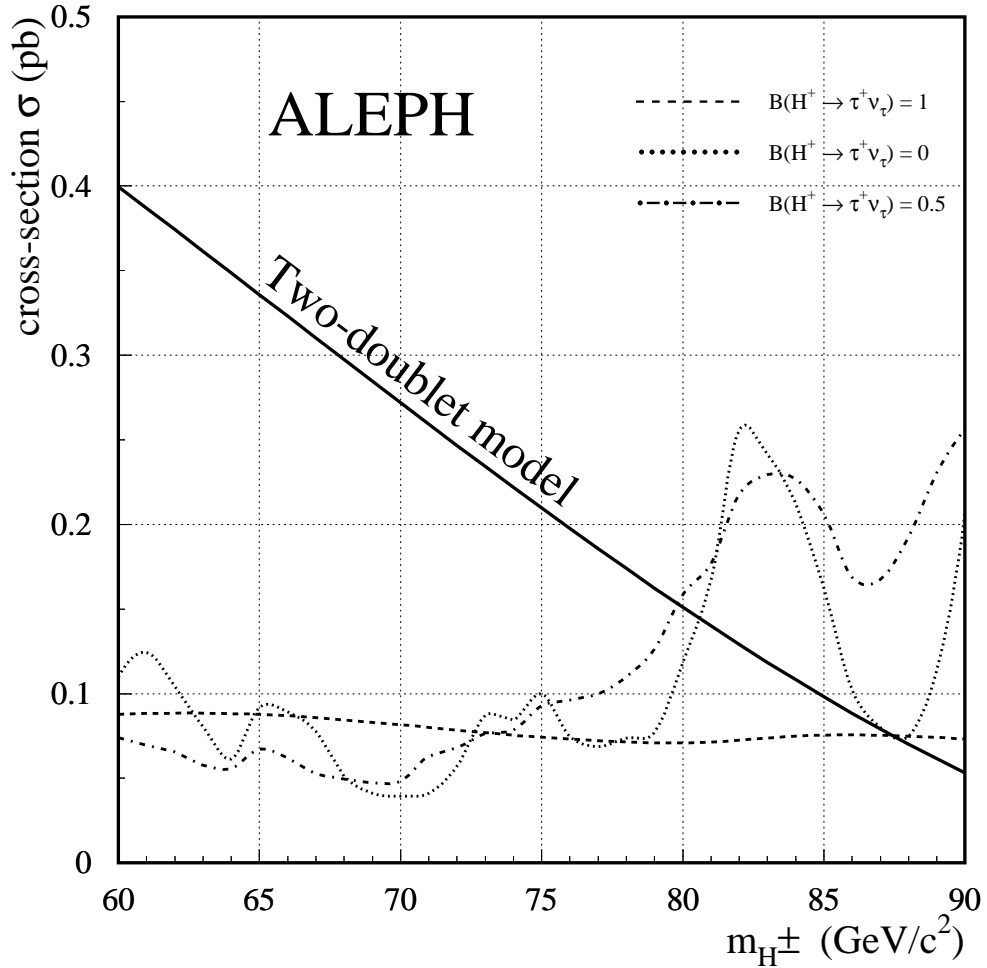}
 \caption{Upper limits at 95\% C.L. on the \hh{} production 
cross section at $\sqrs = 200$~GeV for \btn=1 (dashed line),
\btn=0 (dotted line) and \btn=0.5 (dashed-dotted line).
The charged Higgs boson production cross section in the two-Higgs-doublet
model is shown as a solid curve.}
 \label{xsec_scan}
\end{figure}

Upper limits can also be derived on the \hh{} cross section at
\sqrs{} = 200\,GeV, as a function of the Higgs boson 
mass, for \btn=0, 50 and 100\%.
To combine the data at different centre-of-mass energies,
the limit on the cross section was extrapolated to 200\,GeV
with the expected \sqrs{} dependence for the production of a 
charged scalar particle pair. 
The result is shown in~Fig.~\ref{xsec_scan} as a function
of \mh. 

\newpage

\section{Conclusions}
Pair-produced charged Higgs bosons have been searched for in the three 
final states
\tntn, \tncs{} and \cscs{}, with 629\,\invpb{} of data
collected at centre-of-mass energies from 189 to 209\,GeV. 
No evidence for Higgs boson production
was found and lower limits were set on \mh{} as 
a function of \btn{},
within the framework of two-Higgs-doublet models.
Assuming \btn+\bcs=1, 
charged Higgs bosons with mass below 79.3\,\gevcc{}
are excluded at 95\% C.L., independent of \btn .
\newpage
{\bf {\Large Acknowledgements }}
\vspace{3mm}

It is a pleasure to congratulate our colleagues from the accelerator divisions
for the successful operation of LEP at high energy.
We are indebted to the engineers and technicians in all our institutions
for their contribution to the excellent performance of ALEPH.
Those of us from non-member states wish to thank CERN for its hospitality
and support.

\end{document}